\documentclass[12pt]{iopart}
\usepackage{xcolor}
\usepackage{graphicx}%
\usepackage{dcolumn}%
\usepackage{bm}%
\usepackage{multirow}
\usepackage[numbers,sort&compress,square]{natbib}
\usepackage{acronym}
\usepackage{hyperref}%
\usepackage{verbatim}
\usepackage{comment}
\usepackage[mathlines,running]{lineno}

\graphicspath{{./final_figs/}}

\newcommand{\ts}{\textsuperscript}
\newcommand{\CIERA}{Center for Interdisciplinary Exploration and Research in Astrophysics (CIERA), Northwestern University, 1800 Sherman Ave, Evanston, IL 60201, USA}
\newcommand{\Princeton}{Department of Astrophysical Sciences, Princeton University, 4 Ivy Ln, Princeton, New Jersey, 08544, USA}
\newcommand{\PrincetonPhy}{Department of Physics, Princeton University, Princeton, NJ 08544, USA}
\usepackage{iopams}  
\newcommand{\spinningSingularLowerChia}{\ensuremath{0.02^{+0.05}_{-0.02}}}
\newcommand{\spinningSingularUpperChia}{\ensuremath{0.62^{+0.15}_{-0.10}}}
\newcommand{\spinningSingularPeakChia}{\ensuremath{0.15^{+0.09}_{-0.11}}}
\newcommand{\spinningSingularLowerChib}{\ensuremath{0.00^{+0.01}_{-0.00}}}
\newcommand{\spinningSingularUpperChib}{\ensuremath{0.36^{+0.08}_{-0.07}}}
\newcommand{\spinningSingularPeakChib}{\ensuremath{0.01^{+0.09}_{-0.01}}}
\newcommand{\spinningNonsingularLowerChia}{\ensuremath{0.04^{+0.04}_{-0.02}}}
\newcommand{\spinningNonsingularUpperChia}{\ensuremath{0.59^{+0.10}_{-0.08}}}
\newcommand{\spinningNonsingularPeakChia}{\ensuremath{0.18^{+0.07}_{-0.06}}}
\newcommand{\spinningNonsingularLowerChib}{\ensuremath{0.00^{+0.01}_{-0.00}}}
\newcommand{\spinningNonsingularUpperChib}{\ensuremath{0.35^{+0.07}_{-0.06}}}
\newcommand{\spinningNonsingularPeakChib}{\ensuremath{0.04^{+0.07}_{-0.04}}}
\newcommand{\spinningTruncnormLowerChia}{\ensuremath{0.03^{+0.03}_{-0.01}}}
\newcommand{\spinningTruncnormUpperChia}{\ensuremath{0.57^{+0.14}_{-0.13}}}
\newcommand{\spinningTruncnormPeakChia}{\ensuremath{0.21^{+0.05}_{-0.04}}}
\newcommand{\spinningTruncnormLowerChib}{\ensuremath{0.00^{+0.00}_{-0.00}}}
\newcommand{\spinningTruncnormUpperChib}{\ensuremath{0.35^{+0.08}_{-0.07}}}
\newcommand{\spinningTruncnormPeakChib}{\ensuremath{0.00^{+0.12}_{-0.00}}}
\newcommand{\nonspinningSingularLowerChia}{\ensuremath{0.00^{+0.01}_{-0.00}}}
\newcommand{\nonspinningSingularUpperChia}{\ensuremath{0.40^{+0.08}_{-0.04}}}
\newcommand{\nonspinningSingularPeakChia}{\ensuremath{0.03^{+0.04}_{-0.02}}}
\newcommand{\nonspinningSingularLowerChib}{\ensuremath{0.00^{+0.00}_{-0.00}}}
\newcommand{\nonspinningSingularUpperChib}{\ensuremath{0.18^{+0.05}_{-0.03}}}
\newcommand{\nonspinningSingularPeakChib}{\ensuremath{0.00^{+0.00}_{-0.00}}}
\newcommand{\nonspinningNonsingularLowerChia}{\ensuremath{0.01^{+0.01}_{-0.00}}}
\newcommand{\nonspinningNonsingularUpperChia}{\ensuremath{0.44^{+0.05}_{-0.02}}}
\newcommand{\nonspinningNonsingularPeakChia}{\ensuremath{0.09^{+0.03}_{-0.01}}}
\newcommand{\nonspinningNonsingularLowerChib}{\ensuremath{0.00^{+0.00}_{-0.00}}}
\newcommand{\nonspinningNonsingularUpperChib}{\ensuremath{0.23^{+0.04}_{-0.02}}}
\newcommand{\nonspinningNonsingularPeakChib}{\ensuremath{0.00^{+0.02}_{-0.00}}}
\newcommand{\nonspinningTruncnormLowerChia}{\ensuremath{0.01^{+0.01}_{-0.00}}}
\newcommand{\nonspinningTruncnormUpperChia}{\ensuremath{0.20^{+0.08}_{-0.04}}}
\newcommand{\nonspinningTruncnormPeakChia}{\ensuremath{0.07^{+0.04}_{-0.02}}}
\newcommand{\nonspinningTruncnormLowerChib}{\ensuremath{0.00^{+0.00}_{-0.00}}}
\newcommand{\nonspinningTruncnormUpperChib}{\ensuremath{0.12^{+0.05}_{-0.03}}}
\newcommand{\nonspinningTruncnormPeakChib}{\ensuremath{0.00^{+0.03}_{-0.00}}}
\newcommand{\GWTCLowerChia}{\ensuremath{0.06^{+0.09}_{-0.03}}}
\newcommand{\GWTCUpperChia}{\ensuremath{0.73^{+0.15}_{-0.17}}}
\newcommand{\GWTCPeakChia}{\ensuremath{0.29^{+0.13}_{-0.10}}}
\newcommand{\GWTCLowerChib}{\ensuremath{0.01^{+0.05}_{-0.01}}}
\newcommand{\GWTCUpperChib}{\ensuremath{0.48^{+0.15}_{-0.13}}}
\newcommand{\GWTCPeakChib}{\ensuremath{0.08^{+0.13}_{-0.07}}}
\newcommand{\ChiBFirstPriorUpper}{\ensuremath{0.36}}
\newcommand{\pvalueOneLow}{\ensuremath{0.40}}
\newcommand{\pvalueTwoLow}{\ensuremath{0.11}}
\newcommand{\pvalueOne}{\ensuremath{2.63\times 10^{-6}}}
\newcommand{\pvalueTwo}{\ensuremath{7.64\times 10^{-2}}}
\begin{document}

\title[Disentangling Spinning and Nonspinning Binary Black Hole Populations]{Disentangling spinning and nonspinning binary black hole populations with spin sorting}

\author{Lillie Szemraj$^{1,2}$ and Sylvia Biscoveanu$^{2,3,4}$\footnote{NASA Einstein Fellow}
}
\address{$^1$\Princeton}
\address{$^2$\CIERA}
\address{$^3$\PrincetonPhy}
\address{$^4$Society of Physicists Interested in Non-aligned Spins (SPINS)\footnote{\url{sites.mit.edu/spins}}}

\ead{sbisco@princeton.edu}
\vspace{10pt}
\begin{indented}
\item[]\today
\end{indented}

\begin{abstract}
The individual component spins of \acfp{BBH} are difficult to resolve using gravitational-wave observations but carry key signatures of the processes shaping their formation and evolution. 
Recent analyses have found conflicting evidence for a sub-population of black holes with negligible spin, but the \textsc{Default} spin magnitude population model used in \acf{LVK} analyses cannot formally accommodate an excess of systems with zero spin.
In this work, we analyze several different simulated \ac{BBH} populations to demonstrate that even in the face of this mismodeling, spinning and nonspinning populations can be reliably distinguished using the \textsc{Default} spin magnitude population model coupled with spin sorting.
While typical analyses sort the binary components by their masses, sorting the components by their spin magnitudes instead offers a complementary view of the properties of individual systems consistent with equal mass and of population-level properties, given binary evolution processes like tidal-spin up that predict asymmetric spin magnitudes among the binary components.
We conclude that current observations of the \ac{BBH} population are inconsistent with a fully nonspinning population, but could be explained by a population with only one spinning black hole per binary or a population with up to \textcolor{black}{$\sim 80\%$} nonspinning sources.
\end{abstract}

\vspace{2pc}
\acrodef{O4}[O4]{fourth observing run}
\acrodef{BBH}[BBH]{binary black hole}
\acrodef{IID}[IID]{independently and identically distributed}
\acrodef{LVK}[LVK]{LIGO-Virgo-KAGRA}
\acrodef{PN}[PN]{post-Newtonian}
\acrodef{SOR}[SOR]{spin-orbit resonance}
\acrodef{RMR}[RMR]{reversed mass ratio}
\acrodef{SMR}[SMR]{standard mass ratio}
\acrodef{SNR}[SNR]{signal-to-noise ratio}
\acrodef{VM}[VM]{von Mises}
\acrodef{KDE}[KDE]{Kernel Denisty Estimate}
\acrodef{MC}[MC]{Monte Carlo}
\acrodef{MCMC}[MCMC]{Markov Chain Monte Carlo}
\acrodef{CDF}[CDF]{cumulative distribution function}
\acrodef{PPD}[PPD]{posterior predictive distribution}

\section{Introduction} 
\label{sec:intro}

The spins of \acfp{BBH} carry signatures of the astrophysical processes shaping their formation and evolution. %
For instance, \acp{BBH} formed via isolated binary evolution are expected to have spins that favor alignment with the orbital angular momentum~\cite{Tutukov:1993, Kalogera:1999tq, Grandclement:2003ck, Postnov:2014tza, Belczynski:2016obo, Mandel:2015qlu, Marchant:2016wow, Rodriguez:2016vmx, Stevenson:2017tfq}. In contrast, binaries formed dynamically in dense stellar environments like globular or nuclear star clusters are predicted to have misaligned spins~\cite{Sigurdsson:1993zrm, Miller:2008yw, Zwart:2010kx, Benacquista:2011kv, Rodriguez:2016vmx}.
In addition to the spin tilt angle, the dimensionless spin magnitudes of \acp{BBH}, $\chi=|\vec{S}c/(GM^{2})|$, provide clues to their formation histories. Black holes formed from direct stellar collapse are predicted to have negligible natal spins due to efficient angular momentum transport in their stellar progenitors~\cite{Fuller:2019sxi}. However, the second-born black hole in an isolated binary can be spun up due to tidal effects in its progenitor star~\cite{Qin:2018vaa, Bavera:2020inc, Bavera:2020uch, Bavera:2021evk}. In dense stellar environments, hierarchical mergers including the remnant of a previous merger can lead to a sub-population of \acp{BBH} with spins around $\chi\sim 0.7$ due to the conservation of angular momentum~\cite{Buonanno:2007sv, Tichy:2008du, Lousto:2009ka, Hofmann:2016yih, Fishbach:2017dwv}. Gravitational-wave measurements of \ac{BBH} component spin magnitudes and tilts can thus shed light on the formation channels contributing to the astrophysical population of these sources.

 Analyses of the gravitational-wave data recorded by the \acf{LVK} detectors~\citep{LIGOScientific:2014pky, VIRGO:2014yos, Aso:2013eba, Somiya:2011np, KAGRA:2020tym} provide limited ability to resolve the component spins of individual \acp{BBH}, as these parameters have a minimal effect on the waveform at leading order~\cite{Damour:2001tu, Racine:2008qv, Ajith:2009bn}. %
Population-level analyses of the \ac{BBH} sources reported in GWTC-3, the latest \ac{LVK} catalog~\cite{KAGRA:2021vkt}, suggest most sources have low spin magnitudes~\citep{KAGRA:2021duu, Biscoveanu:2020are, Tong:2022iws, Mould:2022xeu, Galaudage:2021rkt} and find a preference for more aligned-spin than antialigned systems, though a fully isotropic distribution is not ruled out~\cite{Vitale:2022dpa, Golomb:2022bon, Edelman:2022ydv, Callister:2023tgi}. These results are bolstered by analyses looking at the effective aligned and precessing spin parameters~\cite{Miller:2020zox, Roulet:2021hcu, Callister:2022qwb, Banagiri:2025dxo}, $\chi_{\mathrm{eff}}$ and $\chi_{p}$, which capture leading-order spin effects~\cite{Damour:2001tu, Ajith:2009bn, Ajith:2011ec, Santamaria:2010yb, Purrer:2013ojf, Schmidt:2014iyl} and are thus better measured for individual events than the component spins~\cite[e.g.,][]{Vitale:2016avz, Shaik:2019dym}.

The \textsc{Default} population-level spin model employed by the \ac{LVK} through GWTC-3~\cite{LIGOScientific:2018jsj, LIGOScientific:2020kqk, KAGRA:2021duu} assumes that the component spins are \ac{IID} following a Beta distribution for the magnitudes~\cite{Wysocki:2018mpo} and identically distributed following a mixture model between an isotropic distribution and a truncated Gaussian peaked at $\cos\theta=1$ (aligned spin) for the tilts~\cite{Talbot:2017yur}. While a Beta distribution is a convenient choice for the spin magnitudes because it must peak in the interval $\chi \in [0,1]$, it cannot formally accommodate an excess of systems with negligible spin, $\chi\approx0$. \textcolor{black}{This is because such an excess of nonspinning systems is represented by a delta function, $\delta(\chi=0)$, but the Beta distribution only approaches a delta function as the hyper-parameters governing its shape become infinite.} Previous works~\cite{Kimball:2020opk, Kimball:2020qyd, Roulet:2021hcu, Galaudage:2021rkt, Callister:2022qwb, Tong:2022iws, Adamcewicz:2023szp} have explored model variations designed to encapsulate such an excess either directly in the component spin magnitude distribution or in the distribution of $\chi_{\mathrm{eff}}$. %
They generally find that for GWTC-3, up to 80\% of the \ac{BBH} population can be nonspinning but cannot rule out the hypothesis that the entire population consists of binaries with two spinning black holes. However, these studies have disagreed on the astrophysical interpretation of their results; some claim that nonspinning systems represent a large fraction of the \ac{BBH} population~\cite{Galaudage:2021rkt} while others report no evidence for a nonspinning population~\cite{Callister:2022qwb, Tong:2022iws}. 

The components of a binary system are typically categorized by the masses of the objects, where $\chi_{1}$ ($\chi_{2}$) refers to the spin of the more (less) massive black hole. %
However, the underlying \ac{BBH} population has a preference for equal-mass binaries~\cite{KAGRA:2021duu, Farah:2023swu}, introducing a degeneracy in the identification of the binary components. An alternative parameterization was proposed in \cite{Biscoveanu:2020are}, where the components of the binary are instead categorized by their spin magnitudes (spin sorting), $\chi_j$ for $j \in$ $\{A, B\}$ with $A$ referring to the object with highest spin and $B$ referring to the object with lowest spin, $\chi_A \geq \chi_B$. This parameterization was shown to lead to improved constraints on the component spins for individual binaries consistent with equal mass~\cite{Biscoveanu:2020are} and is also convenient for representing populations where only one black hole is spinning ($\chi_{B} = 0$), as may be expected astrophysically for systems formed via isolated binary evolution or for mergers containing one higher-generation black hole formed dynamically. \textcolor{black}{Alternative approaches to avoiding labeling ambiguities in binary spin constraints include using summary statistics like $\chi_{\mathrm{eff}}$, which is invariant under labeling exchange, semi-supervised machine learning methods to determine the optimal binary sorting scheme on an event-by-event basis~\cite{Gerosa:2024ojv}, and spin sorting in the context of distinguishing slow and recycled neutron stars in binary systems~\cite{Zhu:2020zij}.}

In this work, we seek to determine whether the \textsc{Default} \ac{LVK} spin magnitude model coupled with spin sorting can be used to qualitatively distinguish spinning from nonspinning \acp{BBH} on a population level. %
As described in Section~\ref{sec:methods}, we simulate several different populations of \acp{BBH} detectable during the ongoing \ac{LVK} \ac{O4} to enable comparison of the inferred population-level spin magnitude distributions with the \ac{LVK} results.
 Beginning with nonspinning and singly-spinning populations, we perform hierarchical Bayesian inference and find that there are statistically significant differences in the population-level distributions of $\chi_{A/B}$ that could be used to distinguish these two populations, presenting our results in Section~\ref{sec:single_vs_nospin}. We also hierarchically analyze populations with varying mixture fractions between spinning and nonspinning systems in Section~\ref{sec:mixed}.
 By comparing our results to those inferred using the GWTC-3 data, we find support for the interpretation that the majority of \acp{BBH} have at least one spinning black hole. Finally, we conclude and discuss the implications of our analysis in Section~\ref{sec:conclusions}.

\section{Methods} 
\label{sec:methods}
We generate three independent populations of \acp{BBH} that could be detected by a LIGO Hanford-Livingston detector network at their projected O4 sensitivities~\cite{Aasi:2013wya, O4_psds}: one where both black holes are nonspinning, one where only one black hole in each binary is spinning (singly-spinning), and another population where both black holes are spinning. For all populations, we draw events from a uniform distribution on the mass ratio, $\pi(q) = \mathrm{U(0.25, 1)}$ and a power-law prior on the detector-frame chirp mass, $\pi(\mathcal{M}) \propto \mathcal{M}$$^{-3.5}, \mathcal{M} \in [35, 200]~M_{\odot}$. The prior on the luminosity distance is uniform in source-frame time and comoving volume between 100 and 5000 Mpc. The extrinsic source parameters are drawn from the standard priors used in individual-event parameter estimation~\cite[e.g.,][]{KAGRA:2021vkt}. 

Because our focus in this work is on the spin magnitude distribution, the distributions for the other binary parameters are chosen to be simple but astrophysically reasonable. For example, the chirp mass distribution is chosen for consistency with the distribution implied by the primary mass and mass ratio distributions inferred by the \ac{LVK} using GWTC-3 data~\cite{KAGRA:2021duu}. However, we generate simulated source parameters by drawing directly from this detector-frame chirp mass prior rather than from the source-frame primary mass distribution. This is done to avoid a prior mismatch during the individual-event inference step~\cite{Vitale:2020aaz}, as sampling directly in detector-frame chirp mass improves sampler convergence. The prior is restricted to high-mass sources to ensure that all simulated events can be analyzed using the same data duration during the individual-event parameter estimation step. 

The spin magnitudes of both black holes are set to $\chi_{1/2}=0$ for the nonspinning population. For the singly-spinning population, we assume that the mass-sorted spins are identically but not independently distributed, such that for half the events, $\chi_{1}=0$ and $\chi_{2}$ is drawn from a Beta distribution with $\alpha_{\mathrm{true}}=1.014, \beta_{\mathrm{true}}=3.402$ (chosen based on the spin magnitude distribution inferred using GWTC-3 data~\cite{KAGRA:2021duu}). The distributions are switched for the other half of the events. The resulting distributions for the spin-sorted spin magnitudes are $\pi(\chi_{A}) = \mathrm{Beta}(\chi_{A}; \alpha_{\mathrm{true}}, \beta_{\mathrm{true}}), \pi(\chi_{B}) = \delta(0)$. For the fully spinning population, both black hole spin magnitudes are \ac{IID} following the Beta distribution described above. The spin tilt and azimuthal angles of the spinning black holes are assumed to be isotropically distributed.

\textcolor{black}{We add the astrophysical signal generated using the IMRPhenomPv2 waveform model~\citep{Hannam:2013oca} for each simulated event to frequency-domain Gaussian noise colored by the projected O4 power spectral density for the LIGO Hanford and Livingston detectors~\cite{O4_psds}.} Strain data recorded by the LIGO detectors undergoes a matched-filtering search process to determine whether it likely contains a signal or should be discarded as noise~\cite[e.g.,][]{Allen:2005fk, Usman:2015kfa, Adams:2015ulm, Messick:2016aqy}. To mimic this real detection process in our simulated data, we use the network matched filter \ac{SNR} as a proxy for the detection statistic~\cite{Essick:2023upv}, requiring $\mathrm{SNR}_{\mathrm{net}}^{\mathrm{MF}}\geq 8$ for an event to be considered detected. Our resulting populations consist of 156 events for the nonspinning population, 154 events for the singly-spinning population, and 149 events for the fully spinning population passing the threshold to be further analyzed at the population level. 

For each detected event, we first \textcolor{black}{use the \textsc{Bilby} package~\citep{Ashton:2018jfp, Romero-Shaw:2020owr} to perform Bayesian parameter estimation and obtain posterior samples for the parameters of each individual binary. We use the same waveform model and priors} described above for generating the simulated population during the individual-event inference step, with the exception of the spin magnitudes for which we use uniform distributions, $\chi_{1/2} = \mathrm{U}(0,0.99)$, following \ac{LVK} convention (see Table~\ref{tab:priors} for further prior details).

\begin{table}
\centering
\resizebox{\textwidth}{!}{%
\begin{tabular}{p{5.5cm}p{2cm}p{3cm}p{4cm}}
  \br
  \textbf{Parameter Name} & \textbf{Symbol} & \textbf{Prior Shape} & \textbf{Bounds} \\
  \mr
  Luminosity Distance & $d_L$ & $\propto \frac{1}{1+z}\frac{d V_{c}}{d z}\frac{d z}{d d_{L}}$ & (100 Mpc, 5000 Mpc)\\
  Chirp Mass & $\mathcal{M}$ & $\propto \mathcal{M}^{-3.5}$ & (35 \(\textup{M}_\odot\), 200 \(\textup{M}_\odot\)) \\
  Mass Ratio & \textit{q} & Uniform & (0.25, 1)  \\
  Primary Spin & $\chi_1$ & Uniform & (0, 0.99) \\
  Secondary Spin & $\chi_2$ & Uniform & (0, 0.99) \\
  \mr
  Beta distribution $\alpha$ & $\alpha_\chi$ & Uniform & (0.1, 10) - singular\\
  Beta distribution $\alpha$ & $\alpha_\chi$ & Uniform & (1, 10) - nonsingular\\
  Beta distribution $\beta$ & $\beta_\chi$ & Uniform & (0.1, 10) - singular\\
  Beta distribution $\beta$ & $\beta_\chi$ & Uniform & (1, 10) - nonsingular \\
  Truncated Gaussian mean & $\mu$ & Uniform & (0, 1) \\
  Truncated Gaussian width & $\sigma$ & Uniform & (0.05, 1)\\
  \br
\end{tabular}
}
\caption{\label{tab:priors} Priors on the individual simulated events for both populations in the parameter
estimation step (top) and on the hyper-parameters for population inference (bottom).}
\end{table}

We then perform population inference by fitting the mass-sorted component spins of each population with two different population models---Beta (\ac{LVK} \textsc{Default}) and Truncated Gaussian distributions. Specifically, we obtain posteriors on the hyper-parameters $\Lambda$ common to all of the events in each population, which describe the shape of the population prior, $\pi_{\mathrm{pop}}(\theta | \Lambda)$, where $\{\chi_{1}, \chi_{2}\}\in\theta$ for our analysis. For the Beta distribution model, we seek to constrain the hyper-parameters $\{\alpha_{\chi}, \beta_{\chi}\} \in \Lambda$, where the distribution becomes a Delta function at $\chi=0$ ($\chi=1$) as $\alpha_{\chi}\rightarrow0, \beta_{\chi}\rightarrow\infty$ ($\alpha_{\chi}\rightarrow\infty, \beta_{\chi}\rightarrow0$). Distributions with $\alpha_{\chi} < 1$ ($\beta_{\chi} < 1$) are ``singular'', since $p(\chi)=\infty$ for $\chi=0$ ($\chi=1$). \textcolor{black}{We perform analyses using priors that both exclude and include singular Beta distributions, the former to match the \ac{LVK} convention and the latter to allow for distributions that peak at $\chi=0$, since we are interested in constraining \ac{BBH} populations with nonspinning components.} The truncated Gaussian is parameterized in terms of a mean and width, $\{\mu_{\chi}, \sigma_{\chi}\} \in \Lambda$, with bounds $\chi_{\min}=0, \chi_{\max}=1$. The hyper-parameter priors used for population inference are given in Table~\ref{tab:priors}.

The likelihood, $\mathcal{L}(\{d\} | \Lambda)$, of observing an ensemble of data segments, $\{d\}$ given the hyper-parameters $\Lambda$ is obtained by marginalizing over the individual-event parameters $\theta$ and should be modified to account for selection effects~\cite[e.g.,][]{Loredo:2004nn, Thrane:2018qnx, Mandel:2018hfr, Vitale:2020aaz},
\begin{eqnarray}
    \mathcal{L}(\{d\} | \Lambda) \propto \frac{1}{\alpha^{N_{det}}(\Lambda)}\prod_{i}^{N_{det}}\int \mathcal{L}(d_{i} | \theta)\pi_{\mathrm{pop}}(\theta | \Lambda)d\theta, \label{eq:hyper_like} \\
    \alpha(\Lambda) = \int p_{\mathrm{det}}(\theta)\pi_{\mathrm{pop}}(\theta | \Lambda)d\theta,
    \label{eq:VT}
\end{eqnarray}
where the subscript $i$ indicates an individual event in the population, $\alpha(\Lambda)$ represents the probability of detecting an event drawn from a population with hyper-parameters $\Lambda$, $p_{\mathrm{det}}(\theta)$ is the probability of detecting an individual event with binary parameters $\theta$, and $N_{\mathrm{det}}$ is the number of detected sources included in the analyzed population.

The integral in \ref{eq:VT} is typically computed numerically using \ac{MC} integration over ``found'' sensitivity injections drawn from $p_{\mathrm{draw}}(\theta)$ passed through the same detection pipeline as the real data~\cite{Essick:2023toz},
\begin{eqnarray}
    p_{\mathrm{found}}(\theta) \propto p_{\mathrm{det}}(\theta)p_{\mathrm{draw}}(\theta)\\
    \alpha(\Lambda) \propto \left\langle \frac{\pi_{\mathrm{pop}}(\theta_{j} | \Lambda)}{p_{\mathrm{draw}}(\theta_{j})}\right\rangle_{\theta_{j}\sim p_{\mathrm{found}}(\theta_{j})}
    \label{eq:VT_mc}
\end{eqnarray}
However, this integral struggles with convergence for very narrow distributions like the ones we are attempting to probe in our analysis, necessitating a prohibitively high number of found injections~\cite{Talbot:2023pex}. In \ref{ap:selection_effects}, we show using a simulated injection campaign that the spin magnitudes have a negligible impact on the detectability of a given binary \textcolor{black}{for our simulated populations}, particularly at small spins like the ones we are targeting.
Since the spin magnitudes are the only binary parameters that we fit hierarchically in this analysis, we can assume $p_{\mathrm{det}}(\chi_{1/2})=1$, effectively ignoring selection effects, \textcolor{black}{where the mass-sorted spin magnitudes are denoted by $\chi_{1/2}$.}

The integral in \ref{eq:hyper_like} is also calculated using a \ac{MC} integral over the individual-event posterior samples. While the likelihood uncertainty scales close to quadratically with the number of events in the analyzed population for the selection effects \ac{MC} integral, it scales at most linearly for the \ac{MC} integral over individual-event posterior samples~\cite{Talbot:2023pex}, meaning that convergence issues are less severe for this term. \textcolor{black}{While we do not impose any convergence criteria in our hierarchical likelihood during the stochastic sampling step, we find likelihood variances from this term to be $\sigma^{2}_{\mathrm{MC}} < 1$ for even the nonspinning population recovered with the singular Beta model, below the level where convergence issues might affect results.}

We compute the population-level distributions of the spin-sorted spin magnitudes as a post-processing step using order statistics, which assumes that $\chi_{1/2}$ are \ac{IID}. Using this model, $\chi_{A}$ is the larger of two draws from the mass-sorted $p(\chi_{1/2})$ distribution---the first order statistic---and $\chi_{B}$ the smaller~\cite{Biscoveanu:2020are}:
\begin{eqnarray}
    p(\chi_{A} \textcolor{black}{| \Lambda}) &= 2\, p(\chi_{1/2} | \Lambda)\, \mathrm{CDF}(\chi_{1/2} | \Lambda),\\
    p(\chi_{B} \textcolor{black}{| \Lambda}) &= 2\, p(\chi_{1/2} | \Lambda)\left[1-\mathrm{CDF}(\chi_{1/2}| \Lambda)\right]
\label{eq:spin_sorted}
\end{eqnarray}
 We note that we intentionally mismodel the singly-spinning population during the hierarchical inference step by assuming that $\chi_{1/2}$ are \ac{IID}~\footnote{The spins $\chi_{1/2}$ are identically but not independently distributed for the singly-spinning population, because only one component spin magnitude in each binary can be nonzero.}. This allows us to determine how accurately the true $\chi_{A/B}$ distributions can be recovered for this kind of astrophysically plausible population using the order statistics method currently adopted by the \ac{LVK}.
Finally, we calculate the \acp{PPD} implied by our hyper-parameter inference,
\begin{equation}
\label{eq:ppd}
p_\Lambda(\theta|\{d\}) = \int d\Lambda \, p(\Lambda|\{d\}) 
\, \pi_{\mathrm{pop}}(\theta|\Lambda). 
\end{equation}
The \ac{PPD} represents the updated prior on the parameters $\theta$ given the data $\{d\}$. 
To perform our population inference, we use the gravitational-wave population inference library \textsc{GWPopulation}~\citep{Talbot:2019okv} and the \textsc{Dynesty}~\cite{Speagle:2019ivv} nested sampler for both sampling steps.

\begin{figure*}
\begin{center}
\includegraphics[width = \textwidth]{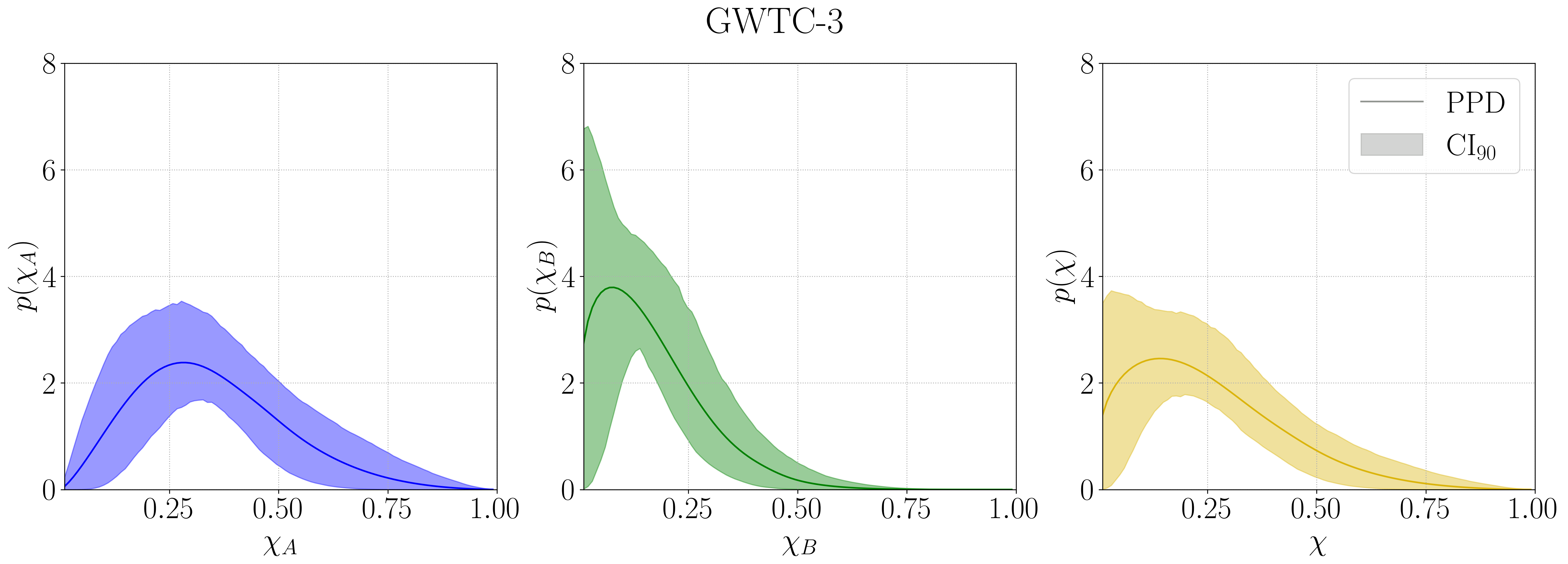}
\caption{\textcolor{black}{\Acfp{PPD}} for $\chi_{A}$ (blue, left), $\chi_{B}$ (green, middle) and $\chi_{1/2}$ (yellow, right) inferred by the \ac{LVK} for the GWTC-3 \ac{BBH} population fit with a Beta distribution (the \textsc{Default} model), shown in the solid lines. The shaded regions correspond to the $90\%$ credible intervals. Figure adapted from Figs. 15, 17 of  \cite{KAGRA:2021duu}.}\label{fig:gwtc3}
\end{center}
\end{figure*}

Figure~\ref{fig:gwtc3} shows the spin magnitude distributions for $\chi_{1/2}$ (yellow) and $\chi_{A/B}$ (blue and green, respectively) inferred by the \ac{LVK} using GWTC-3 data~\cite{KAGRA:2021duu}. While the $\chi_{A}$ distributions peak around $\chi_{A}\sim 0.4$, the $\chi_{B}$ distribution is consistent with peaking at $\chi_{B}=0$, which could suggest a single-spin interpretation for the \ac{BBH} population based on current observations. Below, we provide statistical context for this result by comparing it against our nonspinning and singly-spinning populations. 

\section{Nonspinning vs singly-spinning populations} 
\label{sec:single_vs_nospin}
\begin{figure*}
\begin{center}
\includegraphics[width=0.95\linewidth]{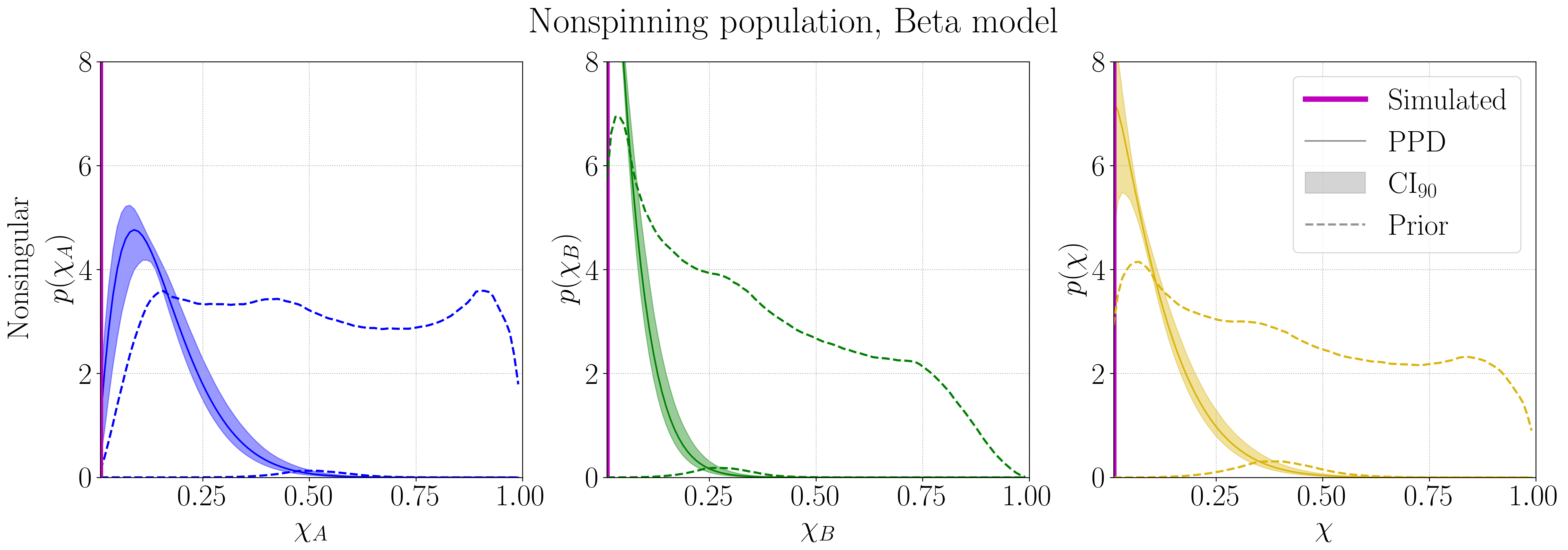}
\includegraphics[width=0.95\linewidth]{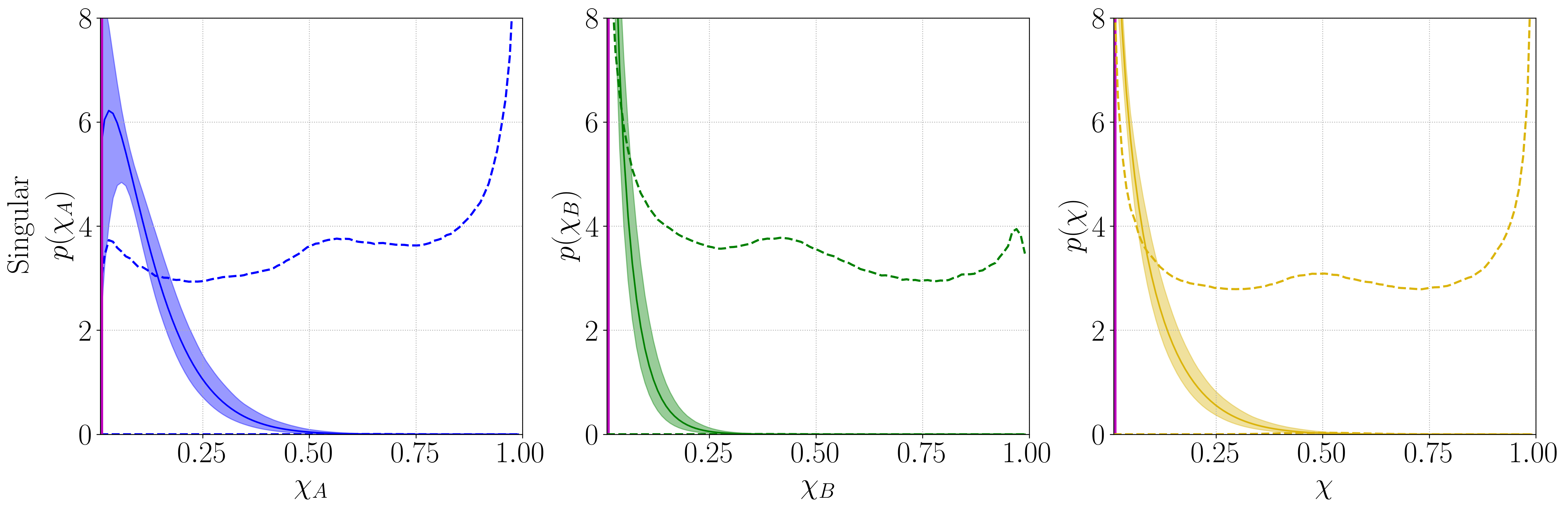}
\caption{Solid lines show the \acp{PPD} for $\chi_{A}$ (blue, left), $\chi_{B}$ (green, middle) and $\chi_{1/2}$ (yellow, right) inferred for the nonspinning \ac{BBH} population when limiting the hyper-parameter prior to nonsingular Beta distributions (top) and including singular Beta distributions (bottom). The shaded regions (dotted lines) correspond to the $90\%$ credible intervals for the posteriors (priors), and the magenta lines show the true distributions.}
\label{fig:nonspinning_beta}
\end{center}
\end{figure*}

\begin{figure*}
\begin{center}
\includegraphics[width = \textwidth]{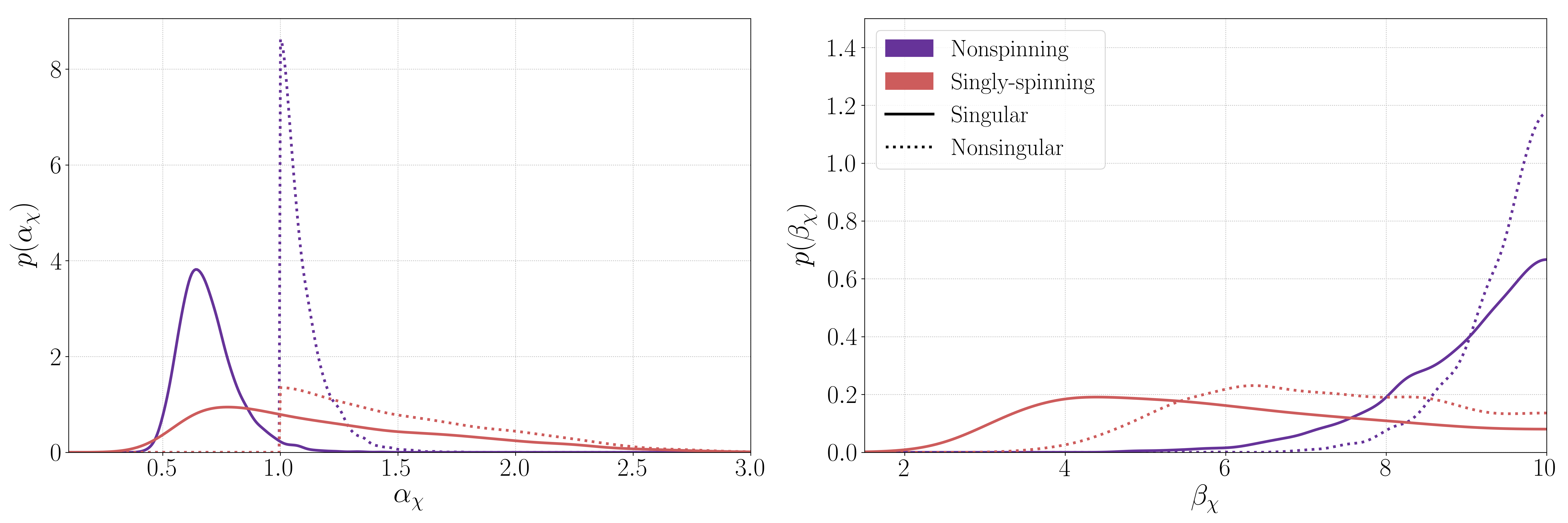}
\caption{Posteriors on the hyper-parameters $\alpha_{\chi}, \beta_{\chi}$ for the nonspinning (spinning) population in purple (red). The solid lines show the results obtained using the priors including singular Beta distributions, and the dotted lines show the nonsingular results.}\label{fig:beta_hyper_kdes}
\end{center}
\end{figure*}

Figure~\ref{fig:nonspinning_beta} shows the spin distributions inferred when analyzing our nonspinning population using the Beta distribution model. As expected, the distributions obtained when allowing for singular Beta distributions in the hyper-parameter prior (bottom panel) are much more strongly peaked towards $\chi=0$ than those obtained when the prior is restricted to nonsingular Beta distributions. As shown in Figure~\ref{fig:beta_hyper_kdes}, for the nonsingular analysis, both the $\alpha_{\chi}$ and $\beta_{\chi}$ posteriors rail against the edges of their priors, indicating that the posteriors are seeking the narrowest $p(\chi)$ distribution allowed by the prior. Because the nonsingular prior limits the allowed values of $\alpha_{\chi}$, the posterior on $\beta_{\chi}$ compensates to produce $p(\chi)$ distributions that are similarly narrow to those obtained in the singular analysis by favoring higher values, comparatively. 
While the posterior on $\beta_{\chi}$ also rails against the upper edge of the prior for the singular analysis, the posterior on $\alpha_{\chi}$ peaks away from the lower bound of the prior. This is likely caused by a combination of the population mismodeling that does not formally allow for delta function distributions and the statistical uncertainty in the measurements of the spins individual events, which retain support for nonzero spins.
This result suggests that hyper-parameter prior railing is not a robust indicator of mismodeling of a predominantly nonspinning \ac{BBH} population using singular Beta distributions as the hierarchical model.

Even though $p(\chi_{A} = 0) = 0$ by definition for the first order statistic, when singular distribution are allowed, this probability drop-off occurs so sharply that the \ac{PPD} and 90\% credible region appear to include distributions with significant nonzero probability at $\chi_{A}=0$. Specifically, we find that the distribution peaks at $\chi_{A} = \nonspinningSingularPeakChia$ for the analysis including singular distributions. Even in the nonsingular case, the $\chi_{A}$ distribution peaks at $\chi_{A} = \nonspinningNonsingularPeakChia$, much lower than the peak inferred at $\chi_{A} = \GWTCPeakChia$ using the GWTC-3 data. As shown in Table~\ref{tab:results}, the 99\textsuperscript{th} percentile of the $\chi_{A}$ distribution inferred for the nonspinning population using the nonsingular Beta distribution model is inconsistent with that inferred using the same model applied to GWTC-3 at $> 90\%$ credibility, suggesting that the GWTC-3 data are inconsistent with a fully nonspinning \ac{BBH} population when analyzed with the \textsc{Default} \ac{LVK} model assuming the mass-sorted spin magnitudes are \ac{IID}. 

\begin{figure*}
\begin{center}
\includegraphics[width=0.95\linewidth]{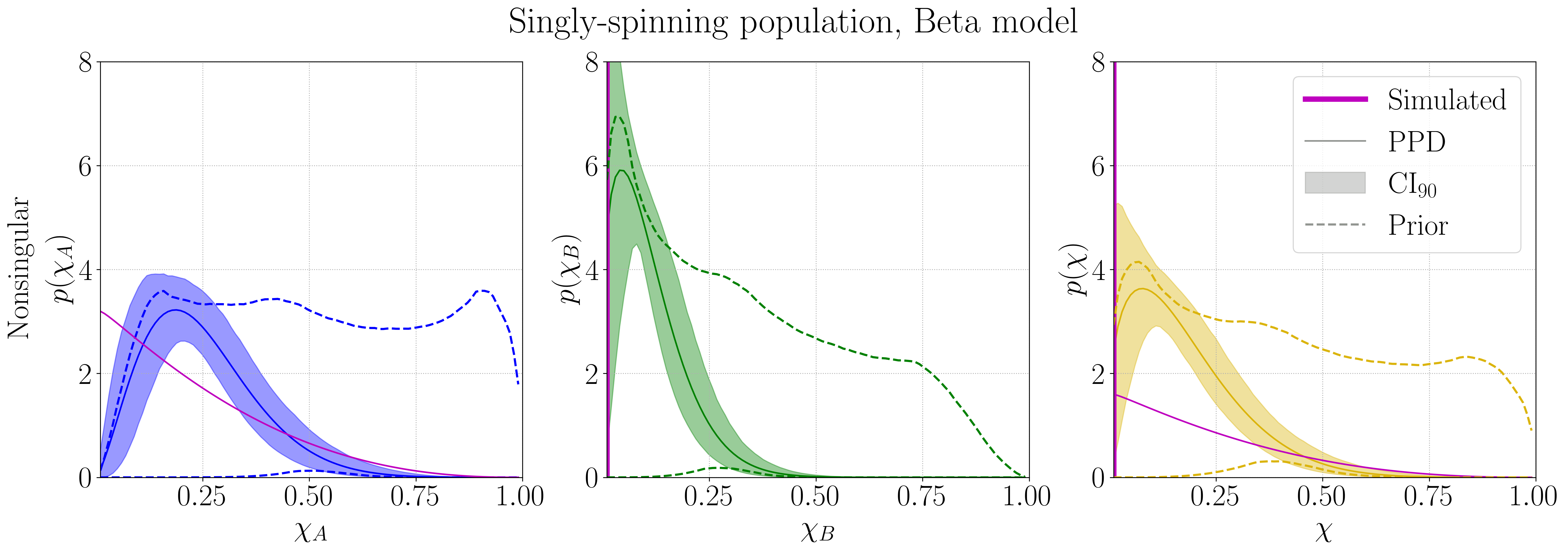}
\includegraphics[width=0.95\linewidth]{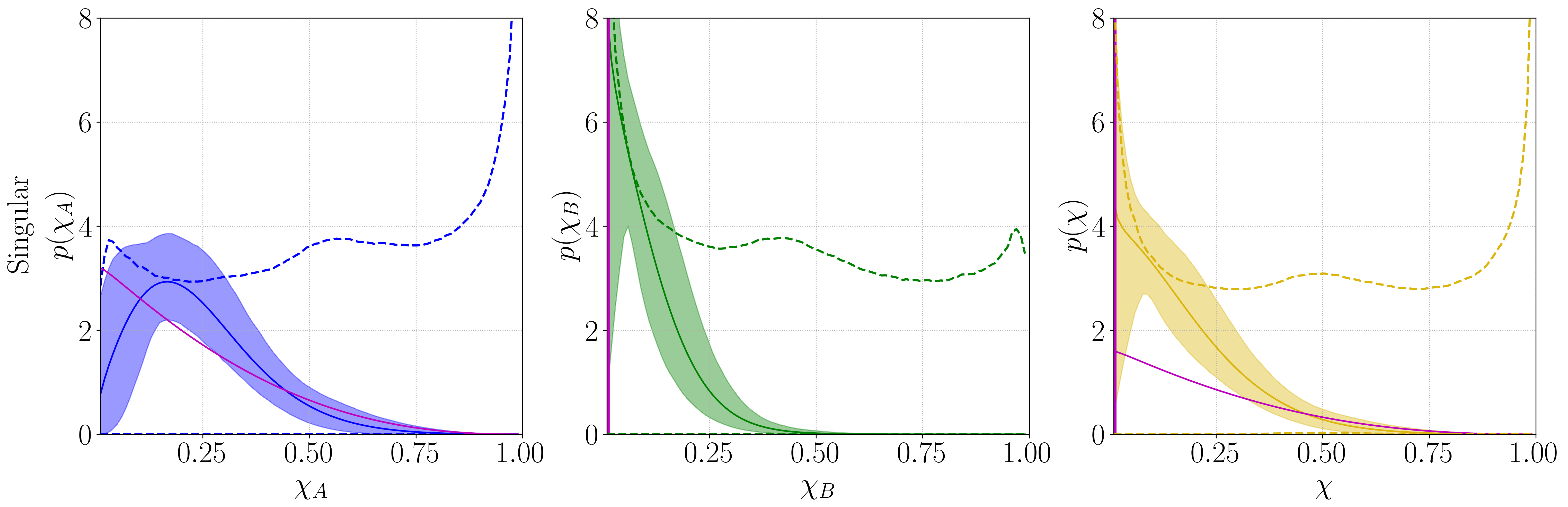}%
\caption{Same as Figure~\ref{fig:nonspinning_beta} but for the singly-spinning population}\label{fig:singlespin_beta}
\end{center}
\end{figure*}

In contrast, the inferred spin magnitude distributions for the singly-spinning population shown in Figure~\ref{fig:singlespin_beta} are much less strongly peaked at $\chi=0$. In this case, we find that the $\chi_{A}$ distribution peaks at $\chi_{A}=\spinningNonsingularPeakChia $ ($\chi_{A}=\spinningSingularPeakChia$) for the nonsingular (singular) case, and $\chi_{A}$ distributions peaking at $\chi_{A}=0$ are ruled out at $> 90\%$ credibility regardless of whether singular Beta distributions are included in the prior. Both inferred $\chi_{B}$ distributions are consistent with peaking at $\chi_{B}=0$, but are less narrowly peaked than for the nonspinning population. Specifically, we find $\chi_{B,99\%} = \spinningNonsingularUpperChib$ for the singly-spinning population when limited to nonsingular Beta distributions compared to $\chi_{B,99\%} = \nonspinningNonsingularUpperChib$ for the nonspinning population. 

While the $\chi_{A}$ distributions that we recover are qualitatively similar to the true Beta distribution used to generate the spins of the individual binaries in our singly-spinning population, the true distribution is not recovered within the 90\% credible region for either the singular or nonsingular Beta model. This is due to twofold mismodeling: 1) The mass-sorted spin population model we have chosen does not allow for a spike at $\chi=0$, meaning that the resulting $\chi$ distribution must somehow split the difference between the half of the population with $\chi=0$ and the other half drawn from a Beta distribution, and 2) The mass-sorted spins of the individual events are not \ac{IID}. If we apply order statistics to the implied \ac{IID} mass-sorted spin distribution used for simulating individual sources, 
\begin{equation}
p(\chi) = \frac{1}{2}\left(\delta(\chi=0) + \mathrm{Beta}(\chi; \alpha_{\mathrm{true}}, \beta_{\mathrm{true}})\right)
\end{equation}
the resulting distributions, $f_{\chi_{A}}(\chi) \neq \mathrm{Beta}(\chi; \alpha_{\mathrm{true}}, \beta_{\mathrm{true}}),\ f_{\chi_{B}}(\chi) \neq \delta(\chi)$. We emphasize that this mismodeling is intentional to gauge whether the differences in the inferred distributions using the current \ac{LVK} \textsc{Default} spin magnitude model allow us to qualitatively distinguish between singly-spinning and nonspinning populations.

\begin{table}
\caption{\label{tab:results}Median and 90\% credible interval on the 1st and 99th percentiles and the peaks of the $\chi_{A}$ and $\chi_{B}$ distributions inferred for different spin populations and spin magnitude models.}
\centering
\resizebox{\textwidth}{!}{%
\begin{tabular}{llcccccc}
\br
Population & Model & \multicolumn{3}{c}{$\chi_{A}$} & \multicolumn{3}{c}{$\chi_{B}$} \\ \cline{3-5} \cline{6-8}
           &       & 1\% & 99\% & Peak & 1\% & 99\% & Peak \\ \ns 
\mr
Spinning & Singular Beta        & \spinningSingularLowerChia & \spinningSingularUpperChia & \spinningSingularPeakChia & \spinningSingularLowerChib & \spinningSingularUpperChib & \spinningSingularPeakChib \\
         & Nonsingular Beta     & \spinningNonsingularLowerChia & \spinningNonsingularUpperChia & \spinningNonsingularPeakChia & \spinningNonsingularLowerChib & \spinningNonsingularUpperChib & \spinningNonsingularPeakChib \\
         & Truncated Gaussian   & \spinningTruncnormLowerChia & \spinningTruncnormUpperChia & \spinningTruncnormPeakChia & \spinningTruncnormLowerChib & \spinningTruncnormUpperChib & \spinningTruncnormPeakChib \\
\mr
Nonspinning & Singular Beta     & \nonspinningSingularLowerChia & \nonspinningSingularUpperChia & \nonspinningSingularPeakChia & \nonspinningSingularLowerChib & \nonspinningSingularUpperChib & \nonspinningSingularPeakChib \\
            & Nonsingular Beta  & \nonspinningNonsingularLowerChia & \nonspinningNonsingularUpperChia & \nonspinningNonsingularPeakChia & \nonspinningNonsingularLowerChib & \nonspinningNonsingularUpperChib & \nonspinningNonsingularPeakChib \\
            & Truncated Gaussian& \nonspinningTruncnormLowerChia & \nonspinningTruncnormUpperChia & \nonspinningTruncnormPeakChia & \nonspinningTruncnormLowerChib & \nonspinningTruncnormUpperChib & \nonspinningTruncnormPeakChib \\
\mr
GWTC-3 & Nonsingular Beta & \GWTCLowerChia & \GWTCUpperChia & \GWTCPeakChia & \GWTCLowerChib & \GWTCUpperChib & \GWTCPeakChib \\
\br
\end{tabular}%
}
\end{table}

\begin{figure*}
\begin{center}
\includegraphics[width=0.95\linewidth]{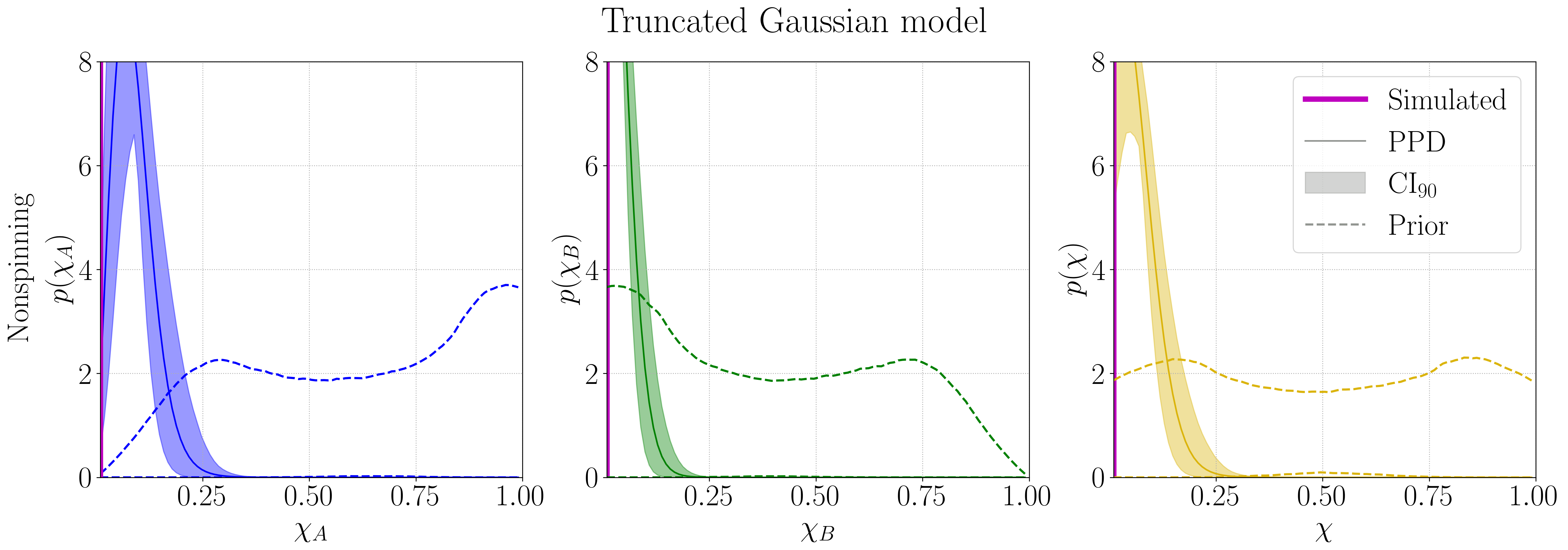}
\includegraphics[width=0.95\linewidth]{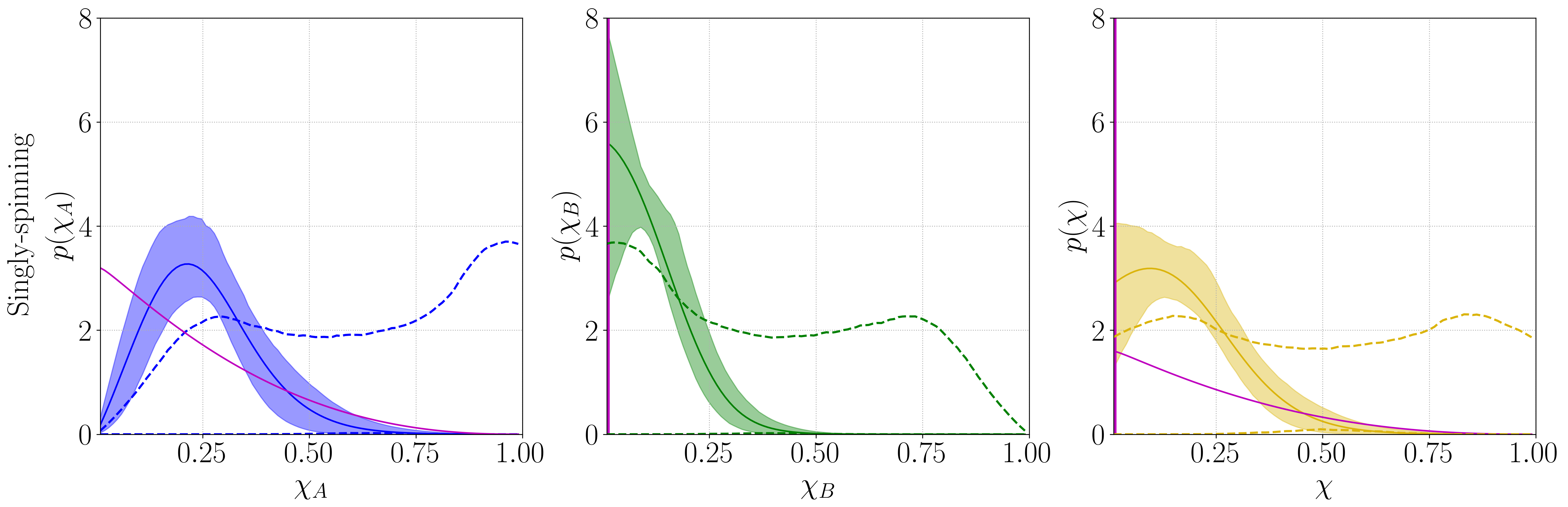}%
\caption{Solid lines show the \acp{PPD} for $\chi_{A}$ (blue, left), $\chi_{B}$ (green, middle) and $\chi_{1/2}$ (yellow, right) inferred for the nonspinning (top) and singly-spinning (bottom) \ac{BBH} populations analyzed using a Truncated Gaussian population model. The shaded regions (dotted lines) correspond to the $90\%$ credible intervals for the posteriors (priors), and the magenta lines show the true distributions.}\label{fig:truncnorm}
\end{center}
\end{figure*}

The Beta distribution is convenient for parameterizing the spin magnitude distribution as it is bounded on the physical domain $\chi \in [0,1]$, but it does not allow for distributions with finite probability support at the edges of this domain. To address this limitation, we also analyze our two simulated populations hierarchically using a Truncated Gaussian population model, which allows for distributions peaking at $\chi=0,1$ without requiring $p(\chi)=\infty$. We note that using the Truncated Gaussian distribution introduces another kind of intentional mismodeling, since the spinning black holes in our singly-spinning population were drawn from a Beta distribution in spin magnitude, which includes asymmetry that may not be captured by a Truncated Gaussian distribution. However, the prior that we choose for the Truncated Gaussian hyper-parameters is flexible enough that it can qualitatively reproduce the behavior of the true simulated $\chi_{A}$ distribution. 

In Figure~\ref{fig:truncnorm}, we show the inferred spin magnitude distributions for both the nonspinning (top) and singly-spinning (bottom) populations recovered with the Truncated Gaussian model. For the nonspinning population, the distributions for all three spin parameters are even narrower using the Truncated Gaussian population model compared to even the nonsingular Beta model, with $\chi_{A, 99\%} = \nonspinningTruncnormUpperChia$ compared to $\chi_{A, 99\%} = \nonspinningSingularUpperChia$. Conversely, the inference for the singly-spinning population is consistent between the two population models used. The $\sigma_{\chi}$ hyper-parameter posterior peaks at the lower edge of its prior for the nonspinning population, indicating that the posteriors are seeking the narrowest possible distribution in this case. For both populations, the posteriors on $\mu_{\chi}$ also peak at the lower edge of the prior, consistent with an excess of systems with negligible spin in both cases.
These results indicate that the degree to which evidence can be recovered for a nonspinning population is limited by both the prior and inherent shape of the parameterized model used for hierarchical inference when delta function distributions cannot be formally accommodated. The Truncated Gaussian distribution is not necessarily better suited to recovering nonspinning or singly-spinning populations than a Beta distribution.

\section{Mixed populations}
\label{sec:mixed}
Most previous works focused on distinguishing spinning and nonspinning sub-populations have directly constrained the fraction of nonspinning systems in the \ac{BBH} population~\cite{Galaudage:2021rkt, Tong:2022iws, Callister:2022qwb, Adamcewicz:2023szp}. While the Beta and Truncated Gaussian models explored here and used by the \ac{LVK} cannot directly constrain this fraction, we can use these models to analyze mixed populations to determine the qualitative properties of the resulting inferred population-level distributions. We generate 11 populations of 149 events each with mixture fractions evenly distributed in $f_{\mathrm{nospin}} \in [0,1]$ by combining our simulated nonspinning and fully spinning populations. We analyze each population with the Beta distribution model including both singular and nonsingular distributions.

\begin{figure*}
\centering
\includegraphics[width = \textwidth]{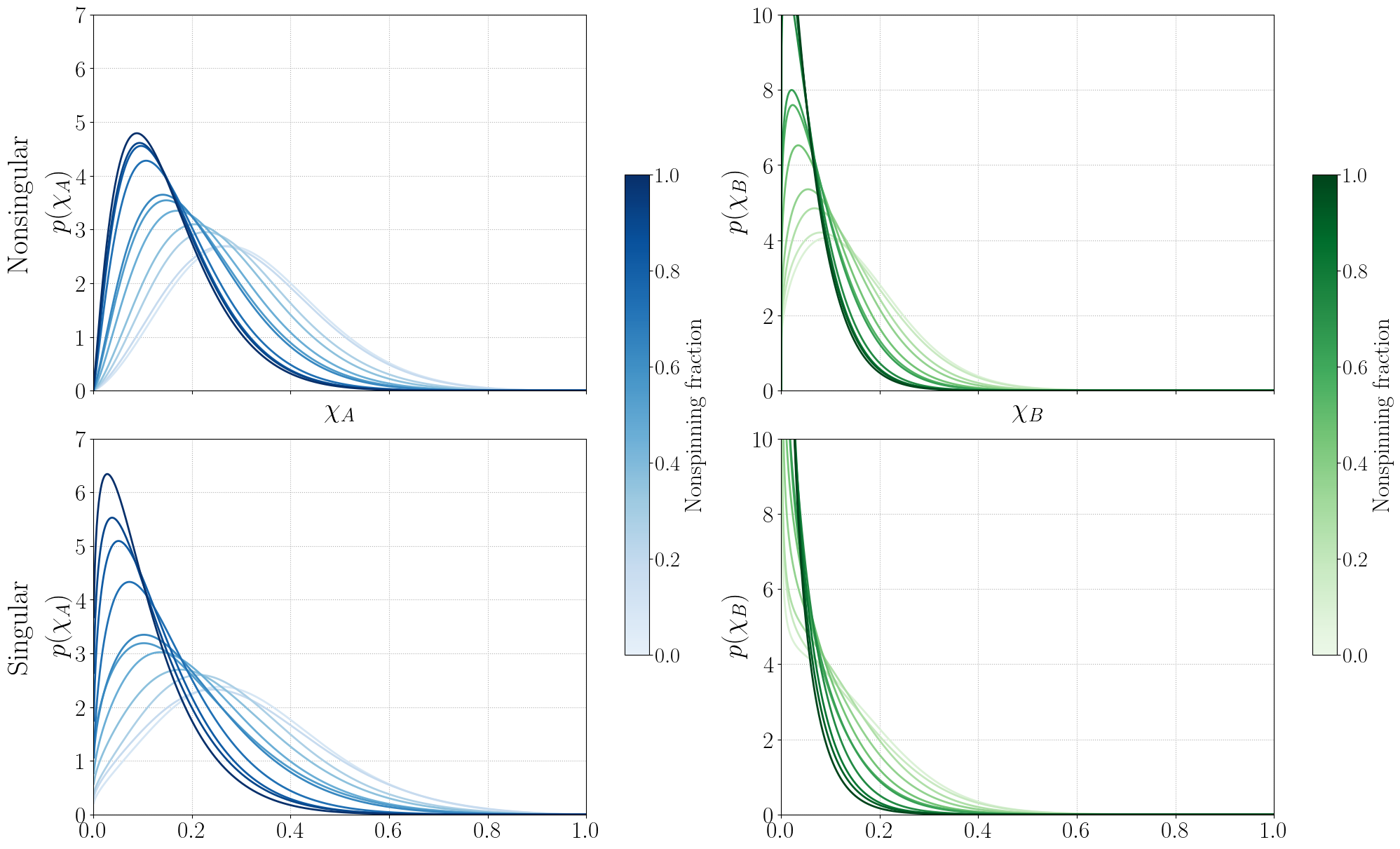}
\caption{PPDs for $\chi_{A}$ (left) and $\chi_{B}$ (right) for eleven different simulated populations with a varying fraction of nonspinning events, ranging from all nonspinning to all spinning. The top (bottom) row shows the results obtained when singular Beta distributions are excluded (included) in the prior.}
\label{fig:mixed_pop_ppds}
\end{figure*}

The \acp{PPD} inferred for $\chi_{A},\ \chi_{B}$ for all 11 populations are shown in Figure~\ref{fig:mixed_pop_ppds}. As expected, both distributions become more strongly peaked towards $\chi=0$ as the nonspinning fraction increases. The differences between the singular and nonsingular results also become more pronounced as the nonspinning fraction increases, as nonsingular distributions peaking at $\chi=0$ can more easily accommodate the increasing excess of nonspinning systems. The posteriors on the $\alpha_{\chi}$ hyper-parameter always peak at the lower edge of the prior for the nonsingular analysis, as expected for both a nonspinning population with $f_{\mathrm{nospin}} = 1$ and a fully spinning population simulated with $\alpha_{\mathrm{true}} = 1.014$. The $\beta_{\chi}$ posteriors peak at the upper edge of the prior for the populations with $f_{\mathrm{nospin}} \gtrsim 0.7$ and favor higher values for the nonsingular analysis compared to the singular analysis, to compensate for the restricted $\alpha_{\chi}$ prior. As we show below, this means that the tails of the inferred $p(\chi_{A/B})$ distributions for a given nonspinning fraction are similar regardless of the prior restriction.

\begin{figure*}
\centering
\includegraphics[width = \textwidth]{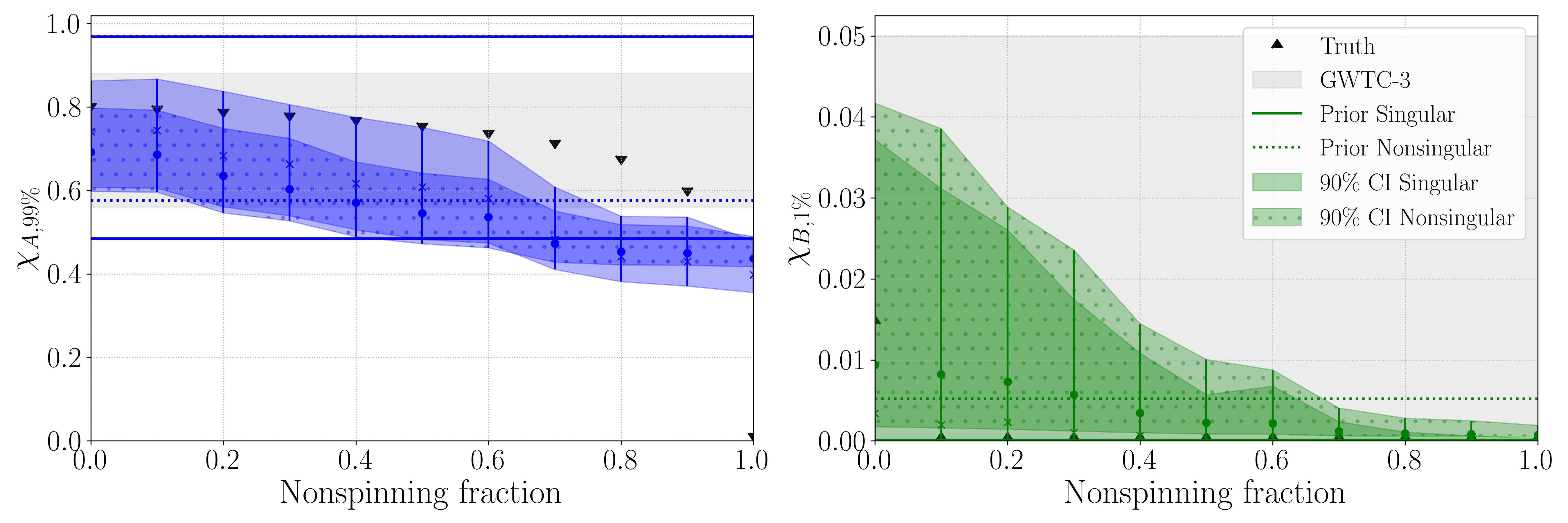}
\caption{99\ts{th} percentile of the inferred $\chi_{A}$ distributions (left) and 1\ts{st} percentile of the inferred $\chi_{B}$ distributions as a function of the fraction of events in the population that are nonspinning. \textcolor{black}{The shaded regions bound the 90\% posterior credible intervals for the singular (solid) and nonsingular (circle-hatched) Beta distribution analyses. The vertical lines span the 90\% credible interval for the discrete mixture fraction values we simulated, and the black markers show the true values for each mixed population.} The 90\% prior credible intervals are bounded by the solid (dotted) horizontal lines for singular (nonsingular) distributions. The upper prior bound for $\chi_{B}$ is $\chi_{B,1\%}=\ChiBFirstPriorUpper$, beyond the upper y-axis limit.}
\label{fig:mixed_pop_limits}
\end{figure*}

Figure~\ref{fig:mixed_pop_limits} shows the inferred 99\textsuperscript{th} percentile of the $\chi_{A}$ distribution and 1\textsuperscript{st} percentile of the $\chi_{B}$ distribution as a function of nonspinning fraction. The uncertainty on both $\chi_{A,99\%}$ and $\chi_{B,1\%}$ generally decreases as the nonspinning fraction increases, indicating that more homogeneous populations are easier to characterize. The uncertainty is also always smaller for the nonsingular Beta analyses, likely due to the restricted prior range (dotted lines in Figure~\ref{fig:mixed_pop_limits}). The 90\% posterior credible region for $\chi_{B,1\%}$ recovered with the nonsingular Beta distribution model excludes the true value of $\chi_{B,1\%}=0$ for all mixture fractions $f_{\mathrm{nospin}} > 0$ because this value is also excluded from the prior, while the true value is always included in the 90\% credible region when singular distributions are included. Meanwhile, the true values of $\chi_{A, 99\%}$ are increasingly excluded from the 90\% posterior credible interval as the nonspinning fraction increases even for the singular Beta analysis; as a consequence of the failure of the model to formally accommodate the excess of systems with $\chi=0$, the posteriors prefer increasingly narrow distributions disfavoring high spins to capture the increasing fraction of nonspinning systems.

Even though $\chi_{B,1\%}$ is $\sim 6$ times more precisely measured than $\chi_{A,99\%}$, it is not measured precisely enough for the fully spinning population with $f_{\mathrm{nospin}}=0$ to exclude some nonspinning fraction with $\chi_{B,1\%}=0$. Meanwhile, the measurements of $\chi_{A,99\%}$ for the nonspinning and fully spinning populations are mutually exclusive. Specifically, we can distinguish predominantly nonspinning populations with $f_{\mathrm{nospin}} \geq 0.8$ from predominantly spinning populations with $f_{\mathrm{nospin}} \leq 0.1$ at 90\% credibility. The value of $\chi_{A,99\%}$ inferred using GWTC-3 data shown in the gray shading in Figure~\ref{fig:mixed_pop_limits} is inconsistent with the values inferred for mixed populations with $f_{\mathrm{nospin}} \gtrsim 0.8$, further indicating that current observations are inconsistent with a fully nonspinning \ac{BBH} population. Similar to previous works using population models tailored to a nonspinning sub-population~\cite{Galaudage:2021rkt, Callister:2022qwb, Tong:2022iws}, our analysis suggests that up to \textcolor{black}{$\sim 80\%$} of \ac{BBH} systems may be nonspinning based on GWTC-3 observations but that current data are not inconsistent with a fully spinning population.

\section{Conclusions}
\label{sec:conclusions}
\textcolor{black}{Ten years after the first direct detection of gravitational waves~\cite{LIGOScientific:2016aoc, LIGOScientific:2016vlm}, our understanding of the \ac{BBH} spin distribution has changed significantly. Before the first direct detection, stellar-mass black holes were expected to be maximally spinning based on observations of galactic black hole--X-ray binaries~\cite{Reynolds:2020jwt}. Now with a catalog of over 300 candidate \ac{BBH} events~\cite{GraceDB}, their spin magnitude distribution is found to favor small spins with a potentially significant nonspinning fraction~\cite{Galaudage:2021rkt, Tong:2022iws, Callister:2022qwb, Adamcewicz:2023szp}. This indicates that either observational or astrophysical selection effects imprint different signatures in the spin distributions of these two classes of black hole binary systems~\cite{Fishbach:2021xqi}. Therefore, accurately determining the nonspinning fraction of \ac{BBH} systems is critical to disentangling the astrophysical processes that shape their evolution.}

In this work, we have shown that the \textsc{Default} \ac{LVK} spin tilt model coupled with spin sorting offers a conceptually simple and robust way to distinguish nonspinning, singly-spinning, and fully-spinning \ac{BBH} populations. %
This model, which assumes the mass-sorted spin magnitudes are \acf{IID} following a Beta distribution, cannot formally accommodate an excess of systems with $\chi=0$ but provides a simple phenomenological description of the spin magnitude distribution that does not require additional individual-event parameter estimation under a nonspinning prior~\cite[e.g.,][]{Galaudage:2021rkt, Tong:2022iws, Adamcewicz:2023szp}. Under this model, the spin-sorted spin magnitudes, $\chi_{A/B}$, are assumed to be the larger and smaller of two draws from the inferred $p(\chi)$ distribution (the first and second order statistics). By calculating these distributions for simulated nonspinning, singly-spinning, and mixed populations and comparing them to the results from GWTC-3, we gain statistical context for previous results claiming that a significant fraction of the \ac{BBH} population could be nonspinning based on current observations.

Even with this intentional mismodeling of a nonspinning (sub)population, we find that the implied distributions of $\chi_{A/B}$ for nonspinning and singly-spinning populations are sufficiently different that they can be distinguished at $> 90\%$ credibility using this model for a population of the size expected by the end of \ac{O4}. These differences are even more statistically significant when the two populations are analyzed with a Truncated Gaussian population model, which relaxes the restriction imposed by the Beta distribution that the probability at the edges of the domain must be either zero or infinite.
We find that the GWTC-3 data are inconsistent with a fully nonspinning poulation at $> 90\%$ credibility. However, they  are consistent with a singly-spinning population, which would be expected for systems formed predominantly via isolated binary evolution that acquire spin in one binary component via tidal spin-up.

We also explore an alternative scenario where only a fraction of the population is nonspinning by analyzing 11 different mixed populations ranging from fully nonspinning to fully spinning (both binary components). Using the 99\textsuperscript{th} percentile as a proxy, we find that the $\chi_{A}$ distributions inferred for predominantly spinning populations with mixture fraction $f_{\mathrm{nospin}} \lesssim 0.1$ are distinguishable from predominantly nonspinning populations with $f_{\mathrm{nospin}} \gtrsim 0.8$ at $> 90\%$ credibility. Using the same summary statistic, we conclude that the GWTC-3 results are consistent with a mixed population where up to \textcolor{black}{$\sim 80\%$} of systems are nonspinning but that a fully spinning population cannot be ruled out, in agreement with previous results obtained with more complex models~\cite{Galaudage:2021rkt, Callister:2022qwb, Tong:2022iws}. While these more complex models allow for direct, quantitative constraints on the fraction of nonspinning systems in the \ac{BBH} populations, our results highlight the continuing utility of the simpler, \ac{LVK} \textsc{Default} model that allows for similar qualitative conclusions to be drawn about the population of \ac{BBH} spins.

While we do not simultaneously model the population-level mass, redshift, and spin angle distributions along with the spin magnitudes in our analysis, we expect that this should have a minimal effect given that correlations between hyper-parameters corresponding to different binary parameters are generally not significant. However, the simultaneous hierarchical modeling of multiple binary parameters would likely increase the likelihood uncertainty from \ac{MC} integration, requiring more careful enforcement of convergence criteria.
In the future, we will also pursue independent fitting of the $\chi_{A/B}$ distributions, breaking the assumption that the mass-sorted spins are \ac{IID}. This will allow us to look for astrophysically-motivated features in the spin magnitude distribution, like an excess in the $\chi_{A}$ distribution at $\chi_{A}=0.7$ from hierchical mergers, or the distribution of $\chi_{A}$ predicted from tidal spin-up in isolated binaries.

\ack
\label{sec:acknowledgments}

This material is based upon work supported by the National Science Foundation under Grant No. AST-2149425, a Research Experiences for Undergraduates (REU) grant awarded to CIERA at Northwestern University and by NSF's LIGO Laboratory which is a major facility fully funded by the National Science Foundation. %
S.B. is supported by NASA through the NASA Hubble Fellowship grant HST-HF2-51524.001-A awarded by the Space Telescope Science Institute, which is operated by the Association of Universities for Research in Astronomy, Inc., for NASA, under contract NAS5-26555.
This research has made use of \textsc{Bilby}~\citep{Ashton:2018jfp}, \textsc{Dynesty} ~\citep{Speagle:2019ivv, Skilling:2004pqw, Skilling:2006gxv}, \textsc{Scipy}~\cite{Virtanen:2019joe}, and \textsc{GWPopulation}\cite{Talbot:2019okv}.
The authors are grateful to Mike Zevin, Vicky Kalogera, Chase Kimball, and Aaron Geller for additional support in the project and throughout the REU program. They also thank Christian Adamcewicz for helpful comments during internal \ac{LVK} document review.
This work used computing resources at CIERA funded by NSF PHY-1726951 and the Quest high performance computing facility at Northwestern University which is jointly supported by the Office of the Provost, the Office for Research, and Northwestern University Information Technology. This manuscript carries LIGO document number P2500458.

\section*{Data availability}
The data that support the findings of this study are openly available on Zenodo~\cite{szemraj_2026_19040701}.

\appendix
\section{Selection Effects}
\label{ap:selection_effects}
\begin{figure*}
\centering
  \includegraphics[clip,width=0.65\columnwidth]{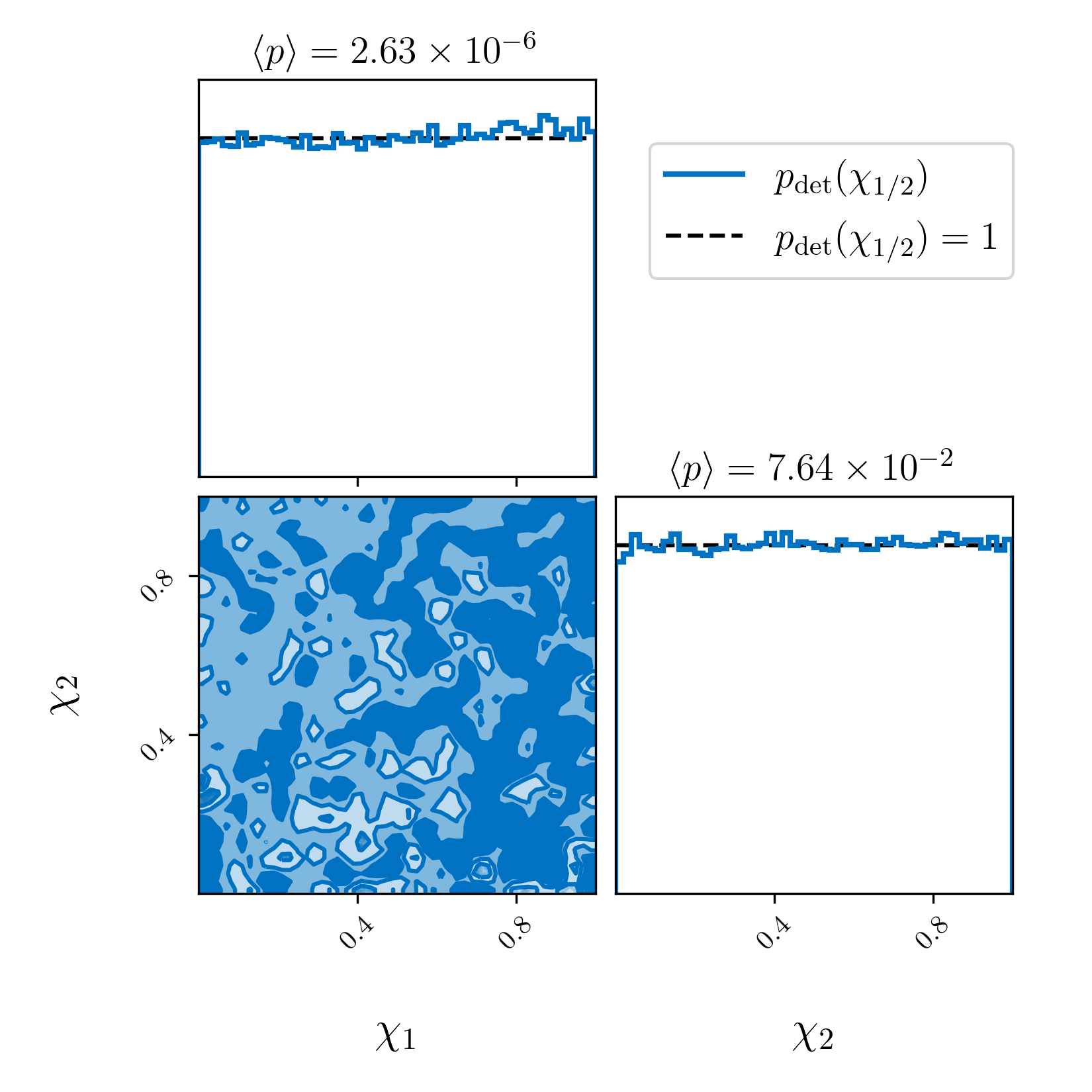}%
\caption{Corner plot of the found sensitivity injections weighted by the inverse of their draw probabilities, indicating the 1 and 2-dimensional detection probabilities as a function of spin magnitude. The shading indicates the 1, 2, and 3-$\sigma$ credible regions, and the dashed black lines are uniform distributions added to facilitate visual comparison. The p-values averaged over 100 Kolmogorov-Smirnov tests of consistency with a uniform distribution for $\chi_{1/2}$ individually are given as the titles for each 1-dimensional panel.}
\label{fig:vt_injections}
\end{figure*}

 To determine the impact of selection effects on the hierarchical likelihood in \ref{eq:VT}, we generate sensitivity injections that meet the same detection criteria as the events included in our various \ac{BBH} populations. The injections are drawn from the same binary parameter priors described in Table~\ref{tab:priors} except for the spin magnitudes, which are drawn from Truncated Gaussian distributions bounded on $\chi \in [0,1]$,
\begin{equation}
    p_{\mathrm{draw}}(\chi_{1},\chi_{2}) = \mathcal{N}_{t}(\chi_{1} ; \mu = 0, \sigma=0.62)\mathcal{N}_{t}(\chi_{2}; \mu=1, \sigma=0.62).
\end{equation}
An event is ``found'' if $\mathrm{SNR}_{\mathrm{net}}^{\mathrm{MF}}\geq 8$, resulting in 191973 injections passing the detection threshold. \textcolor{black}{We note that thresholding on the matched filter \ac{SNR} is an approximation for the sensitivity of a full matched filter search pipeline, where thresholding is typically performed using the false alarm rate~\cite{Essick:2023toz}.} The resulting detection probability as a function of spin magnitude is shown in Fig.~\ref{fig:vt_injections} by plotting the found injections weighted by their inverse draw probabilities, $p_{\mathrm{det}}(\chi_{1/2}) \propto p_{\mathrm{found}}(\chi_{1/2})/p_{\mathrm{draw}}(\chi_{1/2})$. The resulting $p_{\mathrm{det}}(\chi_{2})$ distribution is visually consistent with a uniform distribution (dashed black line), while the $p_{\mathrm{det}}(\chi_{1})$ distribution increases marginally at large values of $\chi_{1}$. This can be explained as a consequence of the orbital hangup effect~\cite{Campanelli:2006uy}, which leads to longer waveforms and hence higher \ac{SNR} for systems with large, aligned spins keeping all other binary parameters fixed. The secondary spin has a subdominant effect on the effective aligned spin that enters the waveform at leading order, so the resulting change in detectability is more apparent in the primary spin distribution~\cite{Ng:2018neg}. 

We perform a Kolmogorov-Smirnov test~\cite{massey1951kolmogorov} to check the consistency of the resulting $p_{\mathrm{det}}(\chi_{1/2})$ with uniform distributions, averaging over 100 different realizations of weighted draws of the Kish effective sample size~\cite{Kish_1965} from $p_{\mathrm{found}}(\chi_{1/2})$. For $\chi_{1}$, we obtain an average p-value of \pvalueOne, indicating that the distribution does indeed deviate from uniformity at large spins. However, when we restrict the distribution to small spin magnitudes, $\chi_{1}<0.5$, which are the focus of this work, the p-value increases to \pvalueOneLow. Neither the the full (\pvalueTwo) nor restricted (\pvalueTwoLow) p-values for $\chi_{2}$ indicate a significant deviation from a uniform distribution. As such, we conclude that ignoring selection effects will have a minimal effect on the inferred spin magnitude distributions for our simulated populations. If anything, we would infer distributions that have marginally more support at high spins rather than an artificial preference for distributions favoring small spins, like those we infer.

\bibliographystyle{iopart-num}
\bibliography{citations.bib}
\end{document}